\documentclass[aps,twocolumn,superscriptaddress,floatfix]{revtex4}
\usepackage[dvips]{graphicx}
\usepackage{latexsym}
\usepackage{amsmath}
\usepackage{amsfonts}
\usepackage{amssymb}
\usepackage{bm}
\usepackage{color}
\usepackage{txfonts}
\begin{document}
	\newcommand{\fig}[2]{\includegraphics[width=#1]{#2}}
	\newcommand{\la}{{\langle}}
	\newcommand{\ra}{{\rangle}}
	\newcommand{\dg}{{\dagger}}
	\newcommand{\upa}{{\uparrow}}
	\newcommand{\dna}{{\downarrow}}
	\newcommand{\ab}{{\alpha\beta}}
	\newcommand{\ias}{{i\alpha\sigma}}
	\newcommand{\ibs}{{i\beta\sigma}}
	\newcommand{\hH}{\hat{H}}
	\newcommand{\hn}{\hat{n}}
	\newcommand{\hc}{{\hat{\chi}}}
	\newcommand{\hU}{{\hat{U}}}
	\newcommand{\hV}{{\hat{V}}}
	\newcommand{\br}{{\bf r}}
	\newcommand{\bk}{{{\bf k}}}
	\newcommand{\bq}{{{\bf q}}}
	\def\gsim{~\rlap{$>$}{\lower 1.0ex\hbox{$\sim$}}}
	\setlength{\unitlength}{1mm}
	\newcommand{{\vhf}}{\chi^\text{v}_f}
	\newcommand{{\vhd}}{\chi^\text{v}_d}
	\newcommand{{\vpd}}{\Delta^\text{v}_d}
	\newcommand{{\ved}}{\epsilon^\text{v}_d}
	\newcommand{{\vved}}{\varepsilon^\text{v}_d}
	\newcommand{{\tr}}{{\rm tr}}
	\newcommand{\pprl}{Phys. Rev. Lett. \ }
	\newcommand{\pprb}{Phys. Rev. {B}}

\title {Chiral flux phase in the Kagome  superconductor AV$_3$Sb$_5$}
\author{Xilin Feng}
\affiliation{Beijing National Laboratory for Condensed Matter Physics and Institute of Physics,
	Chinese Academy of Sciences, Beijing 100190, China}
\affiliation{School of Physical Sciences, University of Chinese Academy of Sciences, Beijing 100190, China}

\author{Kun Jiang}
\email{jiangkun@iphy.ac.cn}
\affiliation{Beijing National Laboratory for Condensed Matter Physics and Institute of Physics,
	Chinese Academy of Sciences, Beijing 100190, China}

\author{Ziqiang Wang}
\affiliation{Department of Physics, Boston College, Chestnut Hill, MA 02467, USA}

\author{Jiangping Hu}
\email{jphu@iphy.ac.cn}
\affiliation{Beijing National Laboratory for Condensed Matter Physics and Institute of Physics,
	Chinese Academy of Sciences, Beijing 100190, China}
\affiliation{Kavli Institute of Theoretical Sciences, University of Chinese Academy of Sciences,
	Beijing 100190, China}
\date{\today}

\begin{abstract}
	We argue that the topological charge density wave phase in the quasi-2D Kagome superconductor AV$_3$Sb$_5$  is a chiral flux phase.   Considering the symmetry of the Kagome lattice, we show that the chiral flux phase has the lowest energy among those states which exhibit  $2\times2$ charge orders  observed experimentally.  This state breaks the time-reversal symmetry and displays anomalous Hall effect. The explicit pattern of  the density of  this state  in real space is calculated.  These results are supported by recent experiments and suggest that these materials are a new platform to investigate the interplay between topology, superconductivity and electron-electron correlations.

\end{abstract}


\maketitle

\section{Introduction}
Owing to its special geometry, materials with a Kagome lattice structure have become  very attractive systems  to investigate  many-body correlation  physics.
For example, the geometry frustration in Kagome magnetism is one of the promising routes towards the quantum spin liquid  \cite{yizhou,qsl1,qsl2}. The electronic structure of  the Kagome lattice also hosts Dirac cones and flat bands, which gives rise to many intriguing quantum phenomena  including nontrivial flat-band responses, tunable Dirac cone etc.  \cite{flatband1,flatband2,flatband3,fqh,ye,yin18}. Additionally,   those Kagome materials also exhibit rich topological physics including anomalous and quantum anomalous Hall, Z$_2$ topological insulator, etc \cite{nagaosa,franz,yin20}. All these fantastic quantum properties make the Kagome lattice being an important platform  to explore  novel physics.

Recently, the newly discovered  quasi-2 dimensional (2D) superconductors AV$_3$Sb$_5$ (A=K,Rb,Cs) may become a very unique system to study the Kagome physics  \cite{ortiz19,ortiz20,ortiz21,lei,shiyanli,hqyuan,jgcheng}.  Up to date, the superconductivity of AV$_3$Sb$_5$  shows   nodal features   \cite{shiyanli,zywang,hezhao}. The study of Josephson junction also provides evidence for possible spin-triplet supercurrent \cite{edgecurrent}.  Besides the superconducting ground state, a topological $2\times2$ charge density wave state (CDW) was also observed in KV$_3$Sb$_5$ by scanning tunneling microscopy (STM) under magnetic field  \cite{topocdw}, which may hold the key to explain the observed anomalous Hall effect  \cite{hall20,xhchen} and topological edge state \cite{edgecurrent}.

These experimental results suggest that   AV$_3$Sb$_5$  are intriguing Kagome materials.  However, these experimental findings still lack theoretical understanding. In this work, we will focus on the  CDW order observed in the normal state. We  show that a chiral flux phase (CFP), which provides a  $2\times2$ topological CDW order,  is a natural candidate to explain observed experimental results. The phase is the  lowest energy state that satisfies the symmetry constraints of the Kagome lattice.   The CFP phase breaks the time reversal symmetry  ${\cal T}$ and  gives rise to the quantum anomalous Hall response.

\begin{figure}
	\begin{center}
		\fig{3.4in}{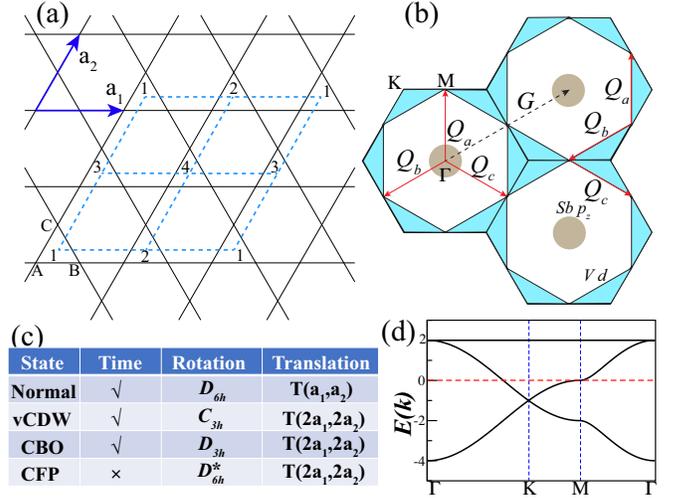}\caption{(a) The Kagome lattice and its translation vector $\mathbf{a}_1$ and $\mathbf{a}_2$. In each unit cell, the sublattice is labeled as A,B,C. The dashed blue lattice is the triangular lattice formed by the unit cell. And the $2\times 2$ unit cell is also plotted. The sub-unit cell for the $2\times 2$  CDW order is also labeled as 1, 2, 3, 4.  (b) The BZ for Kagome lattice. There is one Sb $p_z$ FS (grey circle) around the $\Gamma$ point and the V $d$ FS (blue triangles) around the M point. The M point FS is around the $1/4$ zone filling. The inter-scattering vectors between M points are indicated by $Q_a$,$Q_b$,$Q_c$. (c) Table for symmetry properties for the normal state and three $2\times2$ symmetry breaking states, including time reversal symmetry ${\cal T}$, point group and translation group. The effective point group $D_{6h}^*$ is equal to $\{C_{6h},C_2'{\cal T}C_{6h}\}$.  (d) The band structure for the nearest neighbor Kagome lattice at $5/4$ filling.
			\label{fig:fig1}}
	\end{center}
	\vskip-0.5cm
\end{figure}

\begin{figure*}
	\begin{center}
		\fig{7.0 in}{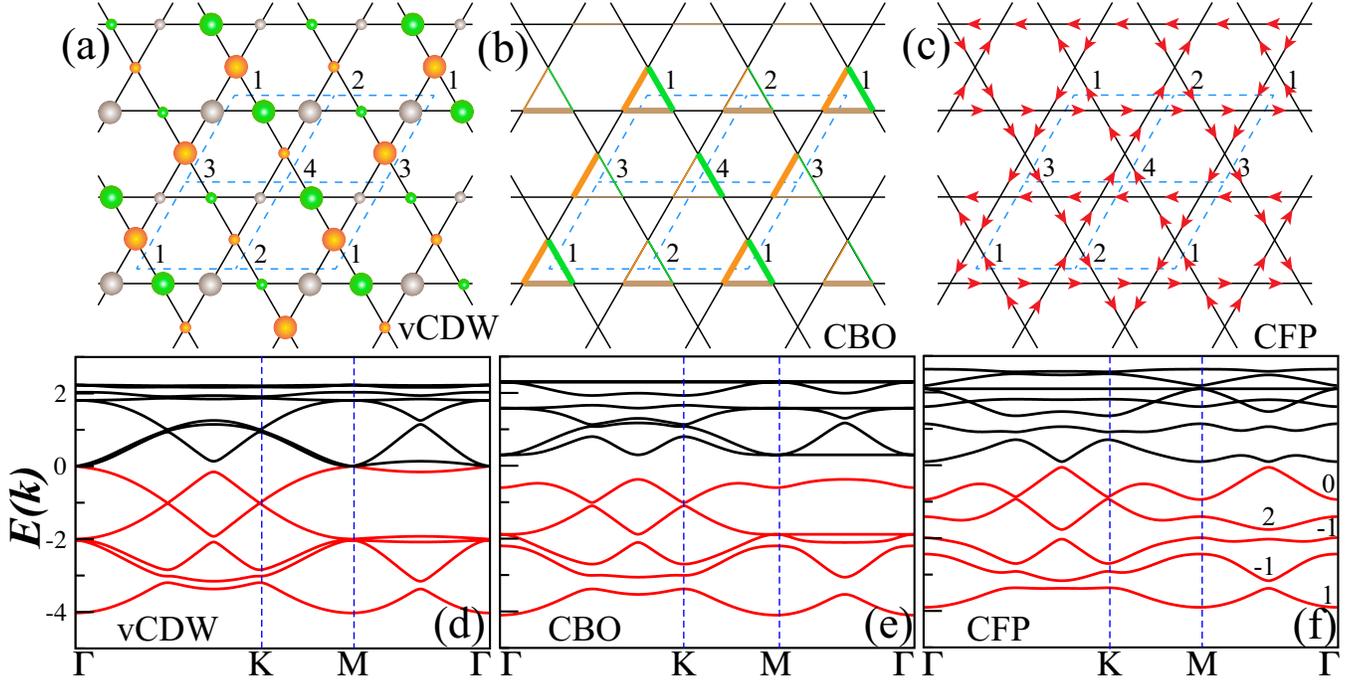}\caption{Three charge orders and their corresponding band structures. (a) the charge configuration for chiral charge density wave (cCDW).   (b) the real bond order configuration for charge bond order (CBO).  (c)  the hopping flux configuration for chiral flux phase (CFP). (d) the band structures for cCDW with $\lambda=0.3$.  (e) the band structures for CBO with $\lambda=0.3$. (f) the band structures for CFP with  $\lambda=0.3$. All the occupied bands are highlight by red for (d), (e), (f). The Chern number for  the occupied bands of CFP is also shown in (f).
			\label{fig:fig2}}
	\end{center}
	\vskip-0.5cm
\end{figure*}

\section{Tight-binding model}
As illustrated in Fig. 1a,  each unit cell of  Kagome lattice contains three sublattices, labeled as A, B, C. And the unit cell forms a triangular lattice with translation vector $\mathbf{a}_1=(1,0)$ and  $\mathbf{a}_2=(\frac{1}{2},\frac{\sqrt{3}}{2})$. This translation group is labeled as $T(\mathbf{a}_1,\mathbf{a}_2)$.  And the point group for the Kagome lattice is $D_{6h}$, as summarized in Fig. 1c.
The $2\times2$ CDW order quadruply enlarges the unit cell, as indicated by the blue dash lattice in Fig. 1a. Each unit cell inside the quadruple order is also labeled as 1, 2, 3, 4. The density functional theory (DFT) calculation shows the normal state of  Kagome AV$_3$Sb$_5$ is a quasi-2D metal, which  is excellently captured by the angle-resolved photoemission spectroscopy (ARPES) measurements  \cite{ortiz20}. Therefore, the AV$_3$Sb$_5$ lies in a weak correlation regime and can be effectively described by a 2D model. Based on DFT and ARPES results,  we find that the in-plane Sb $p_z$ orbital forms one electron pocket around the $\Gamma$ point and the V $d$ orbitals form multiple FSs around the M points, as illustrated in Fig. 1b. Importantly, the V $d$ band FSs touch the van Hove (vH) M points. The inter-scatterings between three vH points give us three important wave vectors $\mathbf{Q}_a=\{0,\frac{2\pi}{\sqrt{3}}\}$, $\mathbf{Q}_b=\{-\pi,-\frac{\pi}{\sqrt{3}}\}$ and $\mathbf{Q}_c=\{\pi,-\frac{\pi}{\sqrt{3}}\}$, which is widely believed  to be the reason for CDW in AV$_3$Sb$_5$. Hence, this $2\times2$ order is also called the 3Q quadruple order.
The STM quasiparticle interference (QPI) results  also show the 3Q scattering between  M points in addition to the  $\Gamma$ point intra-FS scattering  \cite{hezhao,zywang}.  Therefore, to capture the main physics of the topological CDW in AV$_3$Sb$_5$, we can use a minimum  single orbital  model to study  the electronic physics.

The nearest neighbor tight-binding model for Kagome lattice in the basis of $c_{k}=(c_{k,A},c_{k,B},c_{k,C})$  can be written as $H_0=\sum_k c_{k}^\dagger H_k c_{k}$, where
\begin{eqnarray}
	H_k=\begin{bmatrix}
		-\mu & -2t \cos(k_1/2) & -2t \cos(k_2/2) \\
		-2t \cos(k_1/2)  & -\mu & -2t \cos(k_3/2) \\
		-2t \cos(k_2/2) 	&    -2t \cos(k_3/2)  & -\mu
	\end{bmatrix}	
\end{eqnarray}
and  $k_1=k_x$, $k_2=\frac{1}{2}k_x+\frac{\sqrt{3}}{2}k_y$,  and $k_3=-\frac{1}{2}k_x+\frac{\sqrt{3}}{2}k_y$. $\mu$ is the chemical potential and the hopping $t$ is chosen to be 1 as the energy unit.
The band structure for Kagome model is shown in Fig. 1d and the electron filling is tuned to the $5/4$ VH filling. The corresponding FS  agrees with the FS plotted in Fig. 1b.

\section{3Q instability and chiral flux phase}
As described above, the $2\times2$ CDW in AV$_3$Sb$_5$ is widely believed to stem from the multiple Q scatterings between the Kagome VH points, which breaks the translation group down to  $T(2\mathbf{a}_1,2\mathbf{a}_2)$. Actually, these multiple Q FS instability at 3/4 or 1/4 zone filling for hexagonal lattices stems from the seminar discussion in triangular  lattice  \cite{martin}, where the 3Q noncoplanar chiral spin density wave (SDW) order is found to be the leading instability in the Kondo lattice model and the Hubbard model. The chiral SDW order parameter $\mathbf{S}(\mathbf{r})$ can be written as
\begin{eqnarray}
	\mathbf{S}(\mathbf{r})=S(\cos(\mathbf{Q}_a \cdot \mathbf{r}), \cos(\mathbf{Q}_b \cdot  \mathbf{r}),\cos(\mathbf{Q}_c  \cdot  \mathbf{r}))
\end{eqnarray}
where the spin vector is defined as $(s_x,~s_y,~s_z)$ in the three directions. The four spin vectors of this quadrupled SDW order form a tetrahedron AFM at the triangular lattice. Because of the non-zero spin chirality $\mathbf{S}_1\cdot (\mathbf{S}_2\times\mathbf{S}_3)$,  this quadruple order breaks the time-reversal symmetry and causes a quantum anomalous Hall  effect  \cite{martin}.
Besides this chiral SDW order, various FS instabilities, such as  $d+id$ superconductivity, have been widely studied in the honeycomb and Kagome lattices by mean-field theory  \cite{taoli,motome,kun,jxli}, renormalization group (RG)  \cite{chubukov}, functional RG  \cite{qhwang12,thomale12,qhwang13,thomale13}, and density matrix RG \cite{ying} , etc.

For AV$_3$Sb$_5$, the neutron scattering and the muon spin spectroscopy find little evidence for local magnetic moments or long-range magnetic orders  \cite{ortiz19,graf}. Therefore, we will only consider the spontaneous symmetric breaking in the charge channel. In order to construct  a $2\times2$ order, the order parameter must utilize all the three Q scattering vectors,  which implies the order parameter must contain three components in analogy to  the spin $(s_x,~s_y,~s_z)$. The key idea  for $2\times2$  charge order is finding this three-component vector.  Fortunately, the Kagome lattice still has the sublattice degree of freedom beside the spin space.
We can use the three sublattices A, B, C to form a three-component  vector.

Obviously, the first choice is the charge density for each sublattice
\begin{eqnarray}
	\hat{\mathbf{n}}(\mathbf{R})=(\hat{n}_A,\hat{n}_B,\hat{n}_C)
\end{eqnarray}
where the $\mathbf{R}$ is the coordinate for the unit cell formed by the sublattices A, B, C. We name this charge order as the vector charge density wave (vCDW).
The order parameter for this vector can be written as
\begin{eqnarray}
	\boldsymbol{\Delta}_{vCDW}(\mathbf{R})=\lambda (\cos(\mathbf{Q}_a \cdot \mathbf{R}), \cos(\mathbf{Q}_b \cdot  \mathbf{R}),\cos(\mathbf{Q}_c  \cdot  \mathbf{R}))
\end{eqnarray}
where the $\lambda$ is the coupling strength.
Then, the charge order Hamiltonian can be written as
\begin{eqnarray}
	H=H_0-\sum_{R}\boldsymbol{\Delta}_{vCDW}\cdot \hat{\mathbf{n}}(\mathbf{R})
\end{eqnarray}
The charge density for this vCDW order is illustrated in Fig. 2a. The corresponding band structures are also plotted in Fig. 2d. From the band structures in Fig. 2d, we find the vCDW order can not lift the M point degeneracy from the 3Q folding at $\lambda=0.3$. Hence, this state can not be the ground state.

Besides the charge density $\hat{\mathbf{n}}(\mathbf{R})$,  the hopping bonds inside each unit cell can also form a vector
\begin{eqnarray}
	\hat{\mathbf{O}}(\mathbf{R})=(c_A^\dagger c_B,c_B^\dagger c_C,c_C^\dagger c_A)
\end{eqnarray}
More specifically, there are two choices for the hopping bond order parameters, real or imaginary.
If  the order parameter is real, a charge bond order (CBO) is formed, as illustrated in Fig. 2b.
The order parameter for this order can be written as
\begin{eqnarray}
	\boldsymbol{\Delta}_{CBO}(\mathbf{R})=\lambda (\cos(\mathbf{Q}_a \cdot \mathbf{R}), \cos(\mathbf{Q}_b \cdot  \mathbf{R}),\cos(\mathbf{Q}_c  \cdot  \mathbf{R}))
\end{eqnarray}
Then, the charge order Hamiltonian can be written as
\begin{eqnarray}
	H=H_0-\sum_{R}\boldsymbol{\Delta}_{CBO}\cdot \hat{\mathbf{O}}(\mathbf{R})
\end{eqnarray}
The band structure for CBO at $\lambda=0.3$ is shown in Fig. 2b. The CBO opens a large gap at the M point becoming a possible charge order solution. And this CBO is similar to the CBO proposed in Ref.  \cite{thomale13}.

On the other hand, if  the order parameter is imaginary, a chiral flux phase is formed, which can be written as
\begin{eqnarray}
	\boldsymbol{\Delta}_{CFP}(\mathbf{R})=i\lambda (\cos(\mathbf{Q}_a \cdot \mathbf{R}), \cos(\mathbf{Q}_b \cdot  \mathbf{R}),\cos(\mathbf{Q}_c  \cdot  \mathbf{R}))
\end{eqnarray}
As the current must conserve at each lattice site without charge accumulation, we can determine the chiral flux phase, as shown in Fig. 2c. Adding this order parameter and the remaining terms due to current constraint, the band structure for CFP Hamiltonian is obtained in Fig. 2f.  The CFP also opens gap on the FSs at the M point and lifts most of the degeneracy  below the Fermi level. More importantly, the ground state energy for CFP is 0.195t lower than the CBO per unit cell and 0.435t lower than the CDW per unit cell. Therefore, the CFP is the most promising ground state for $5/4$ filled Kagome lattice.

Besides the ground state energy comparison, the most unconventional  behavior for CDW state in  AV$_3$Sb$_5$  is  the chiral anisotropy found by STM measurements under magnetic field  \cite{yin20}. In addition, the anomalous Hall measurements also point out the CDW should break the time-reversal symmetry  \cite{hall20,xhchen}. These results support the CFP as  the ground state compared to the other two  time-reversal  conserved charge orders. To confirm this, we calculate the intrinsic Hall responses using the Kubo formula  \cite{dixiao,tknn,hatsugai}, which is directly related to the Chern number for each band
\begin{eqnarray}
	\sigma_{xy}=\frac{e^2}{\hbar} \sum_{k,n\neq m}\frac{[f(\epsilon_{nk})-f(\epsilon_{mk}) ]Im(v_{x}^{nm}v_{y}^{mn})}{(\epsilon_{nk}-\epsilon_{mk})^2}
\end{eqnarray}
where the $f(x)$ is the Fermi distribution function and $v_{x/y}^{nm}=\langle nk| \hat{v}_{x/y}|mk\rangle$ is the matrix element of the velocity operator. The Chern number for each occupied CFP band is listed in Fig. 2f. The CFP is also an QAH insulator with $C=1$ for each spin sector. The total Hall response is $\sigma_{xy}=2\frac{e^2}{\hbar}$ for the CFP charge order.  However, it is important to note that the AV$_3$Sb$_5$ is a multi-orbital system,  there are Fermi surfaces that are not captured in our simple one-orbital model and can not be fully gapped out by the CFP order.  Therefore, the Hall conductance should not be quantized.

\begin{figure}
	\begin{center}
		\fig{3.4in}{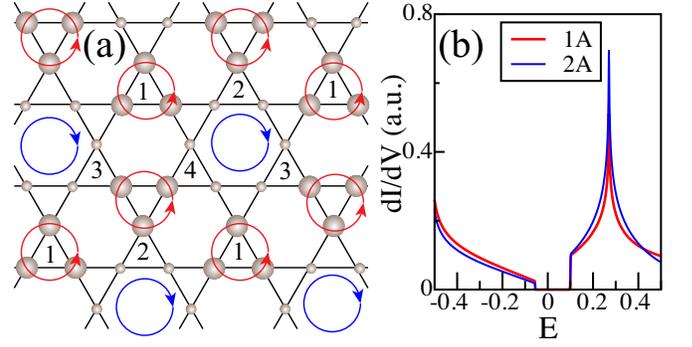}\caption{(a) The relative charge configuration for CFP. The charge density for the anti-clockwise flux triangular sites is larger than the charge density for the clockwise flux hexagonal  sites. (b) The tunneling density of states for site 1A and site 2A.
			\label{fig:fig3}}
	\end{center}
	\vskip-0.5cm
\end{figure}

The charge density distribution for the CFP state at each site is plotted in Fig. 3a. We find that there are two types of charge sites and two special flux loops. As shown in Fig. 3a, the two anti-clockwise triangle current flux loops (red circles)  form a honeycomb lattice and the clockwise hexagonal current flux (blue circle) forms  a triangular lattice. The charge density for each site at the anti-clockwise flux triangular is relatively larger than the charge density at the hexagonal current flux, as indicated in Fig. 3a.  This charge distribution clearly gives rise to a $2\times2$ charge order. Choosing two representative charge sites, we also calculate the tunneling density of states at sites 1A and 2A. They also show distinct features, as plotted in Fig. 3b.

 From the symmetry breaking point of view,  the normal state of  Kagome lattice contains the symmetry $D_{6h}\times  {\cal T}$ in addition to the translation group $T(\mathbf{a}_1,\mathbf{a}_2)$. The $D_{6h}$ is generated by the $C_6$ rotation along the $z$ axis, $C_2'$ along y axis and inversion symmetry ${\cal I}$. As summarized in Fig. 1c, the vCDW and CBO break the $C_6$ rotation down to the $C_{3h}\times  {\cal T}$ and $D_{3h}\times  {\cal T}$ respectively. Interestingly, although CFP breaks the ${\cal T}$ symmetry, there is still one effective point group symmetry $D_{6h}^*$ generated by the $C_6$ , ${\cal I}$ and $C_2' {\cal T}$ . Hence, the CFP  keeps all the point group symmetry of the normal state by doubling the translation vectors. All these symmetry properties can be further examined by  symmetry-selective measurements, like the Kerr effect. In addition,  the coupling between the superconductivity and the CFP  is highly confined by their symmetry characters.  It is an intriguing problem to investigate how this ${\cal T}$ breaking symmetry phase correlated with the superconductivity in AV$_3$Sb$_5$ in the future. And microscopically, an extended Hubbard model with on-site Hubbard interaction U and the nearest-neighbor Coulomb interaction V is needed to stabilize this CFP solution, which will be considered in our following works.

\section{Discussion and summary}
The CFP resembles the loop-current order and  the d-density wave state for the pseudogap phases in cuprates superconductors proposed by C. Varma  and S. Chakravarty  et al.  respectively   \cite{varma1,varma2,varma3, chakravarty}. For loop current order, a current loop is formed inside one  Cu and two O triangle while staggered fluxes are formed in neighboring Cu square plaquettes for the d-density wave state.
Such a loop current or  d-density wave can lead to an  unusual magnetic order  as measured by polarized neutron diffractions  \cite{varma1,yuanli,bourges}.  Thus, we believe that  the polarized neutron diffraction may  find a similar signal for the  CFP order in AV$_3$Sb$_5$.  This signal is related to the orbital magnetism associated with the chiral flux, which is found to be around $0.01\frac{e}{2\hbar}t a^2$  \cite{orbital_magnetsim,dixiao}, where $a$ is the lattice constant and $e$ is the electron charge.

In summary, we discuss the possible $2\times2$ charge order states in  AV$_3$Sb$_5$, including vCDW, CBO and CFP.  We find the CFP state is the lowest energy state, which naturally gives rise to the time-reversal symmetry breaking and anomalous Hall effect. Our findings will provide a new understanding for the topological CDW state of the Kagome material AV$_3$Sb$_5$.

\textbf{Conflict of interest}

The authors declare that they have no conflict of interest.

\textbf{Acknowledgments}

We thank Jiaxin Yin,  Zhenyu Wang, Jinguang Cheng, and Geng Li for useful discussions. This work was supported by the National Basic Research Program of China (2017YFA0303100), the Ministry of Science and Technology of China (2016YFA0302400), the National Natural  Science Foundation of China (NSFC-11888101, NSFC-11674370 and NSFC-11674278), Beijing Municipal Science and Technology Commission Project
(Z181100004218001), the Strategic Priority Research Program of Chinese Academy of Sciences (XDB28000000 and XDB33000000), and the Information Program of the Chinese Academy of Sciences (XXH13506-202). K.J. acknowledges support from the start-up grant of IOP-CAS. Z.W. is supported by the U.S. Department of Energy, Basic Energy Sciences Grant No. DE-FG02-99ER45747.

\textbf{Author contributions}

Xilin Feng, Kun Jiang, Ziqiang Wang and Jiangping Hu jointly identified the problem, performed the calculations and analysis, and wrote
the paper.

\end{document}